\documentclass[letterpaper, 10 pt, conference]{ieeeconf}
\usepackage{booktabs}
\usepackage{graphics}
\usepackage{amsmath}
\usepackage{xcolor}
\usepackage{dsfont}
\usepackage{graphicx}
\usepackage[export]{adjustbox}

\usepackage{stfloats}

\DeclareMathOperator*{\minimize}{minimize}
\renewcommand \baselinestretch{0.98}
\usepackage{fullpage}
\usepackage{amssymb,amsmath}\newcommand{\RR}{\mathbb R}
\usepackage{morefloats}

\usepackage{subcaption}

\usepackage[font=small,labelfont=bf, figurename=Fig.]{caption}

\usepackage{graphicx}

\usepackage{algpseudocode}
\usepackage{algorithm}
\algnewcommand\algorithmicforeach{\textbf{for each}}
\algdef{S}[FOR]{ForEach}[1]{\algorithmicforeach\ #1\ \algorithmicdo}
\algnewcommand\algorithmicinput{\textbf{Input:}}
\algnewcommand\algorithmicoutput{\textbf{Output:}}
\algnewcommand\Input{\item[\algorithmicinput]}
\algnewcommand\Output{\item[\algorithmicoutput]}

\usepackage{setspace}
\usepackage[switch]{lineno} 

\usepackage[backend=biber, style=ieee, citestyle=numeric-comp]{biblatex}
\usepackage{blindtext}
\usepackage[affil-it]{authblk}

\IEEEoverridecommandlockouts

\title{\LARGE \bf
A Greedy Graph Search Algorithm Based on Changepoint Analysis for Automatic QRS Complex Detection 
}

\author[1,*]{Atiyeh Fotoohinasab}
\author[1]{Toby Hocking}
\author[1]{Fatemeh Afghah}

\affil[1]{School of Informatics, Computing and Cyber Systems at Northern Arizona University}
\affil[*]{Corresponding author: Atiyeh Fotoohinasab, af2329@nau.edu}

\renewcommand{\baselinestretch}{.92}
\begin{document}

\maketitle
\thispagestyle{empty}
\pagestyle{empty}

\onecolumn
\begin{abstract}
The electrocardiogram (ECG) signal is the most widely used non-invasive tool for the investigation of cardiovascular diseases. Automatic delineation of ECG fiducial points, in particular the R-peak, serves as the basis for ECG processing and analysis. This study proposes a new method of ECG signal analysis by introducing a new class of graphical models based on optimal changepoint detection models, named the graph-constrained changepoint detection (GCCD) model. The GCCD model treats fiducial points delineation in the non-stationary ECG signal as a changepoint detection problem. The proposed model exploits the sparsity of changepoints to detect abrupt changes within the ECG signal; thereby, the R-peak detection task can be relaxed from any preprocessing step. In this novel approach, prior biological knowledge about the expected sequence of changes is incorporated into the model using the constraint graph, which can be defined manually or automatically. First, we define the constraint graph manually; then, we present a graph learning algorithm that can search for an optimal graph in a greedy scheme.
Finally, we compare the manually defined graphs and learned graphs in terms of graph structure and detection accuracy. We evaluate the performance of the algorithm using the MIT-BIH Arrhythmia Database. The proposed model achieves an overall sensitivity of 99.64\%, positive predictivity of 99.71\%, and detection error rate of 0.19 for the manually defined constraint graph and overall sensitivity of 99.76\%, positive predictivity of 99.68\%, and detection error rate of 0.55 for the automatic learning constraint graph.
\end{abstract}


\twocolumn
\section{INTRODUCTION}
\label{intro}
The electrocardiogram (ECG) is a quasi-periodic biomedical signal that provides information about cardiac muscle electrical activities. One cardiac cycle in a typical ECG signal is delineated by arrangements of P, the QRS complex, T waves, and PQ and ST segments. Correct R-peak detection is the first and most critical step in almost all ECG analysis methods. The R-peak is the highest and only positive peak within the QRS complex, reflecting the ventricular depolarization of the heart's electrical activity. Precise detection of the R-peak location plays a critical role in obtaining the morphology of the QRS complex and revealing the location of other ECG fiducial points. Furthermore, R-peak localization serves as the basis for automated determination of the heart rate, which is a significant criterion for heart arrhythmia diagnoses such as premature atrial contraction, tachycardia, and bradycardia. Many other diseases can also be diagnosed in a non-invasive way using R-peak detection due to the relationship between heart rate variability and several physiological systems (e.g., vasomotor, respiratory, central nervous, and thermoregulatory).

Various approaches have been proposed in the literature for detecting R-peaks in an ECG signal \cite{beraza2017comparative}. Typically, these methods consist of two main steps: pre-processing and detection. In the pre-processing step, the algorithm attempts to eliminate the noise and artifacts and to highlight the relevant sections of the ECG \cite{castells2015simple, sharma2017qrs}. In the second step, various methods are used to locate R-peaks based on the result of the pre-processing step, and then other waves are detected by defining a set of heuristic rules \cite{hou2018real}. However, these approaches suffer from some critical drawbacks that limit their performance in practical applications. First, in real-time data processing and ambulatory care settings, where the collected data are highly noisy, preprocessing-based algorithms are less effective. Second, these algorithms can fail to detect R-peaks in some determinant morphological patterns resulting from certain life-threatening heart arrhythmias due to the time-varying morphology of the QRS complex. Incorrect detection of R-peaks can affect the correct identification of subsequent waves.  
 
The R-peak detection step can be generally accomplished either by implementing a threshold-based technique or by employing an independent threshold technique. The amplitude of the peak and time duration between two consecutive R peaks (i.e., the RR interval) are typically used to determine a suitable threshold \cite{song2015new}. A constant threshold is only efficient for detecting R-peaks within records with normal morphological patterns. Therefore, recent studies have employed adaptive thresholds, for which there is no need to determine the threshold experimentally. In \cite{nayak2019optimally} and \cite{sahoo2016noising}, 
the Hilbert transform with an adaptive thresholding technique was utilized to detect R-peaks. Some threshold-based techniques with other criteria have also been used to specify the threshold. In \cite{hou2018real}, an adaptive threshold concerning the geometric angle between two consecutive samples of the ECG signal was defined. The performance of the threshold-based technique is highly dependent on the selection of initial parameters; hence, it can lead to a significantly higher number of false beats. Therefore, independent threshold techniques are more desirable than the threshold-based technique. 

Most of the state-of-the-art methods for R-peak detection are based on wavelet transform \cite{park2017r, rakshit2017efficient, thirrunavukkarasu2019detection}, simple mathematical operations \cite{nayak2019optimally, PanTompkins1985, gutierrez2015novel}, hidden Markov models, and machine learning. Wavelet transform is a suitable approach for considering the non-stationary behavior of the ECG signal. However, considering the various shapes of the QRS complex, it is difficult to select the optimal mother wavelet or find the required threshold in the detection step of the wavelet transform. Additionally, discrete wavelet transform fails to provide reliable results in a short-recording duration. Mathematical operation-based algorithms have a low computational cost, which is more appropriate for real-time applications and large dataset analysis. However, achieving high performance when the signal-to-noise ratio is high remains challenging for these algorithms. Hidden Markov models are also widely used in ECG segmentation because they are powerful tools for considering the temporal dependency among the waveforms \cite{akhbari2016ecg, monroy2020hidden, huque2019hmm}. The majority of the studies on machine learning-based methods have utilized sparse signal processing to represent an approximation of the nonlinear ECG signal using sparsity constraints \cite{ning2013ecg,zifan2006automated, parviziomran2019optimization, tavakoli2013decoding, monti2018adaptive, tibshirani1996regression}. Some studies have also applied deep learning techniques to detect the ECG waveforms considering its high performance in various classification tasks \cite{xiang2018automatic, tavakoli2017learning}. However, the caveat with deep learning-based approaches is that they need large-scale datasets for the training phase and often suffer from the imbalanced class problem \cite{8683140, mousavi2020single}.

In this paper, we propose a new class of graphical models based on optimal changepoint detection models, named the graph-constrained changepoint detection (GCCD) model, to locate R-peaks in the ECG signal. A changepoint detection model identifies abrupt changes in data when a property of the time series changes. In the non-stationary ECG signal, ECG waves can also be considered as abrupt up or down changes over time during the heart cycle. We exploit the model introduced by Hocking et al. \cite{hocking2017log, hocking2018generalized}, in which a graph-based optimal changepoint detection model was used for detecting abrupt changes in the genomics data. In their work, they propose a new class of functional pruning algorithms with log-linear time complexity in the amount of data, which is capable of handling the large datasets that are common to ECG analysis.

Only a few studies in the literature have applied changepoint detection models for cardiac analysis. Gold et al. \cite{gold2018doubly} adopted a changepoint detection method based on Bayesian inference to extract the onset of the QRS complex over a small time window containing just one QRS complex. In \cite{qi2014novel}, a changepoint detection approach based on the Haar wavelet and Kolmogorov-Smirnov statistic was applied to find normal and abnormal ECG segments within the assembled ECG samples from different ECG datasets. Sinn et al. \cite{sinn2012detecting} analyzed heart rate changes in ECG recordings by detecting abrupt changes in the ordinal pattern distributions, which are used to represent the order structure of a time series. Some studies have also applied changepoint detection models to investigate sleep problems by analyzing heart rate variability in the ECG signal during sleep \cite{staudacher2005new, ma2020multiple}.

To the best of our knowledge, this is the first study in which changepoint detection models have been proposed to detect ECG fiducial points in long records of ECG signals. In this novel framework, prior biological knowledge about the expected sequence of changes can be specified in a constraint graph. Then, functional pruning dynamic programming algorithms can compute the globally optimal model (mean, changes, and hidden states) in fast log-linear time. We furthermore propose a new algorithm for learning the graph structure using labeled ECG data. Therefore, the main contributions of this study are:

\begin{itemize}
\item
A new class of graphical models based on optimal changepoint detection models to detect R-peak positions in the ECG signal. The proposed method does not require any noise removal preprocessing step as it uses the sparsity of changepoints to detect abrupt changes.
\item
A new algorithm to learn the graph structure and parameters using labeled ECG data. Thus, the model's performance is no longer dependent on an expert to encode prior knowledge into the constraint graph. 
\item
Comparison of the learned graphs with the manually constructed graphs in terms of graph structure and detection accuracy. Results demonstrate that there can exist different optimum graph structures for one subject, and the proposed graph learning algorithm can find global optima depending on the initial graph structure.
\end{itemize}

The rest of the paper is organized as follows. In the next section, we describe the proposed model for R-peak detection in the ECG signal. We explain the GCCD model in Section \ref{sec: Graph Constrained Changepoint Detection Model} and the constraint graph in Section \ref{sec:Graph Constraint}.  Section \ref{sec:Graph Constraint} also defines the manual graph and the proposed graph learning algorithm. Section \ref{sec:Experimental Studies} provides a description of the dataset used in this study and a discussion of the results as well as a comparison between the 
performance of the manually defined graphs and learned graphs. Finally, Section \ref{sec:CONCLUSIONS and DISCUSSION} summarizes this research work and its contributions.

\section{Methodology}
 \label{sec:GCCD}
 
\begin{figure*}[htb]
      \centering
      \includegraphics[width=\linewidth,height=\textheight,keepaspectratio]{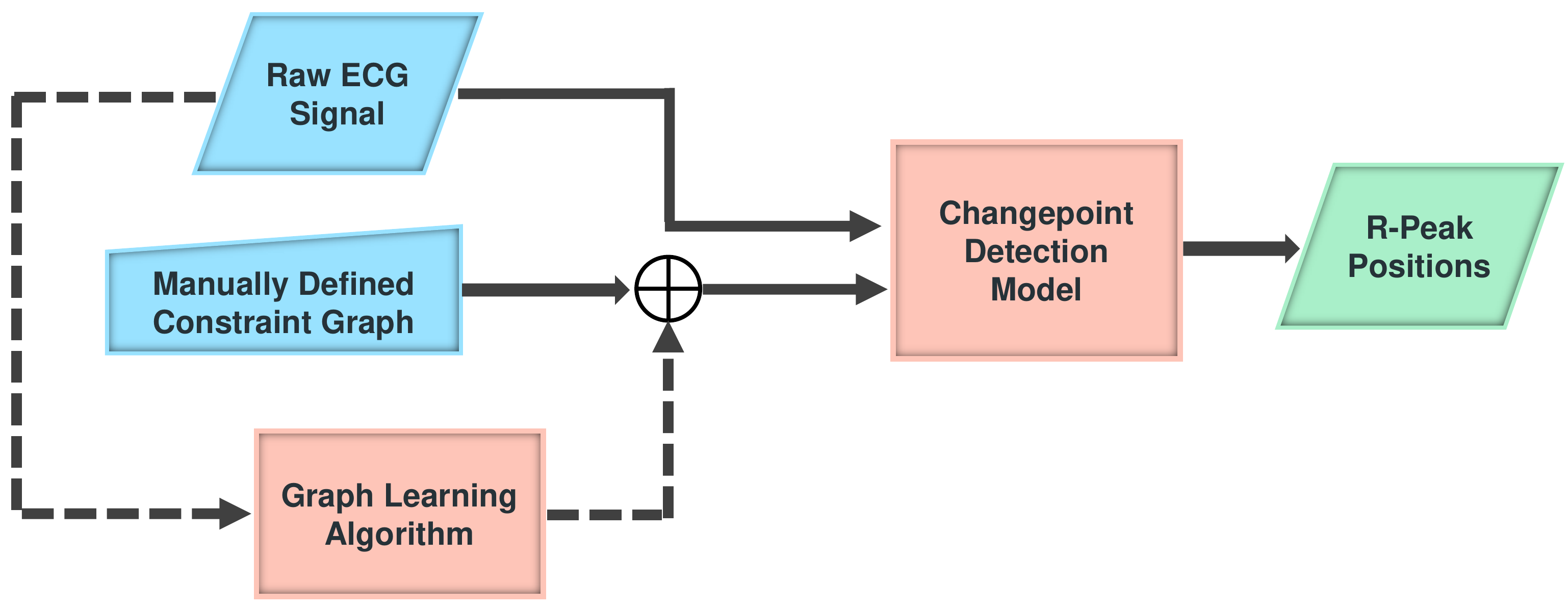}
      \caption{An overview of the GCCD model. The GCCD model takes a constraint graph and a raw ECG signal as inputs and then detects segments corresponding to the nodes of the constraint graph at the output.}
      
      \vspace{-15pt}
      \label{fig:System_model}
   \end{figure*}

The proposed method treats ECG wave detection as a changepoint detection problem for a non-stationary ECG signal. It extracts the R-peaks in the raw ECG signal by representing the periodic non-stationary ECG signal as a piecewise locally stationary time series with constant mean values (i.e., each piece is the mean of one segment of datapoints). The model takes a raw ECG signal and a constraint graph as inputs and computes the onset/offset and the mean of desired segments (i.e., hidden states). Then, the center of each state is associated with the location of a peak. The constraint graph allows the incorporation of prior knowledge into the model and regularizes the model. Figure \ref{fig:System_model} illustrates an overview of the proposed algorithm in the detection of R-peak positions in the ECG signal. It is worth re-emphasizing that the model takes the raw ECG signal as the input, without applying any preprocessing step, as it leverages the sparsity of changepoints to denoise the signal and to detect abrupt changes.

The constraint graph, which encodes the expected sequence of changes in the ECG signal, can be defined manually by an expert or automatically from the data. In the following sections, we describe the details of various parts of the proposed model.

\subsection{Graph-Constrained Changepoint Detection Model}
\label{sec: Graph Constrained Changepoint Detection Model}

ECG fiducial points detection can be defined as the problem of finding abrupt changes over one cardiac cycle caused by changes in statistical characteristics. From this point of view, a proper changepoint detection algorithm can be employed to detect ECG waves in a fast and effective way. We applied the optimal changepoint detection model introduced in \cite{hocking2017log} to localize R-peak positions in the ECG signal. In this model, prior biological knowledge about the expected sequence of changes is incorporated into the model as a graph constraint. Then, a dynamic programming algorithm using functional pruning computes the globally optimal model (mean, changes, and hidden states) in fast log-linear $O(N\log N)$ time.  

We assumed a directed graph $G=(V,E)$ as the constraint graph, where the vertex set $V\in\{1,\dots,|V|\}$ represents the hidden states/segments (not necessarily a waveform), and the edge set $E\in\{1,\dots,|E|\}$ represents the expected changes between the states/segments. Each edge $e\in E$ incorporates the following associated prior knowledge about the expected sequences of changes:

 \begin{itemize}
    \item The source $\underline v_e\in V$ and target $\overline v_e\in V$ are vertices/states for a changepoint $e$ from $\underline v_e$ to $\overline v_e$.
    \item A non-negative penalty constant $\lambda_e\in\RR_+$ is the cost of changepoint $e$.
    \item A constraint function $g_e:\RR\times\RR\rightarrow\RR$ defines the possible mean values before and after each changepoint $e$. If $m_i$ is the mean before the changepoint and $m_{i+1}$ is the mean after the changepoint, then the constraint is $g_e(m_i,m_{i+1})\leq 0$. These functions can be used to constrain the direction (up or down) and/or the magnitude of the change (greater/less than a certain amount).
\end{itemize}

Mathematically, given the input signal $Y=\{y_1,\dots,y_n\}$ and the directed graph $G=(V, E)$, the problem of finding changepoints $c$, segment means $m$, and hidden states $s$ can be solved using the following optimization problem:

\begin{align}
  \label{eq:op-c}
  & \minimize_{
  \substack{
   m\in\RR^N,\, s\in V^N\\
   c\in\{0,1,\dots,|E|\}^{N-1}\\
  }
  } \ \ 
      \sum_{i=1}^N \ell(m_i, z_i) + \sum_{i=1}^{N-1} \lambda_{c_i} \\
        \text{s. t.\ \ } &\ \text{no change: }c_i = 0 \Rightarrow m_i=m_{i+1} ~\& ~s_i=s_{i+1} \, \label{eq:nochange-constraint}\\ \nonumber
    &\ \text{change: }c_i \neq 0 \Rightarrow g_{c_i}(m_i,m_{i+1})\leq 0 ~~\& \\
        &(s_i,s_{i+1})=(\underline v_{c_i},\overline v_{c_i}).\label{eq:change-constraint}
\end{align}
The changepoints $c_i$ can be assigned to any of the pre-defined edges ($c_i\in\{1,\dots,|E|\}$). Consequently, $c_i=0$ indicates no change with zero cost, $\lambda_0=0$. Function (\ref{eq:op-c}) consists of a data-fitting term $\ell$ and a model complexity term $\lambda_{c_i}$ \cite{amini2019iterative, amini2019iterative2}. $\ell$ represents the negative log-likelihood of each datapoint, and $\lambda_{c_i}$ is a non-negative penalty on each changepoint. In other words, $\lambda$ regularizes the number of predicted changepoints/segments by the model so that a larger $\lambda$ reduces the number of changepoints by estimating a more sparse changepoint vector. 
The constraint function $g_e$ also encodes the expected up/down change and the least amplitude gap between the mean of two states. When there is no change $c_i=0$, Constraint~(\ref{eq:nochange-constraint}) forces the model to stay in the current state $s_i=s_{i+1}$ with no change in mean $m_i=m_{i+1}$. However, when there is a change $c_i\neq 0$, Constraint~(\ref{eq:change-constraint}) imposes a change in the mean implied by the constraint function $g_{c_i}(m_i,m_{i+1})\leq 0$ as well as a change in the state $(s_i,s_{i+1})=(\underline v_{c_i},\overline v_{c_i})$. An open-source implementation of the Generalized Functional Prunining Optimal Partitioning (GFPOP) algorithm is available in C++ code inside an R package named GFPOP on GitHub \cite{GFPOP}.
\subsection{Constraint Graph}
\label{sec:Graph Constraint}

The constraint graph $G=(V, E)$ in the optimization problem of Equation (\ref{eq:op-c}) encodes prior biological knowledge about the expected sequences of changes within one cardiac cycle.
It can be designed manually by an expert or be learned from the data by the model. The two following subsections detail both the manual and learning-based designs.

\subsubsection{Manual Graph Definition}
\label{sec: Manual graph definition}

To manually define the constraint graph $G$, we took into account the possible morphological categories for the ECG waves (i.e., P, QRS, and T waves) and the overall morphological properties of the signal in each record \cite{fotoohinasab2020graph}. An expected hidden state/segment in the signal is characterized as a node in the constraint graph, and the required conditions for transition between states are encoded in the edges. The required conditions are determined based on the expected minimum amplitude difference of two successive states and the polarity of each transition (i.e., up/down). 

The caveat with the manual definition of the constraint graph is that it can be inefficient for ECG signal analysis considering the various morphological patterns for each waveform. Furthermore, the model's performance depends on the expert knowledge encoded into the constraint graph. In the next subsection, we explain the proposed graph learning algorithm for learning the constraint graph using the R-peak labels provided by the gold standard.

\subsubsection{Constraint Graph Learning}
\label{sec:Graph Learning}

To automate the R-peak detection task, we modified the previous model by learning the constraint graph from the data (see the dashed part in Figure \ref{fig:System_model}). In this new framework, the proposed model takes the raw signal and an initial graph structure as inputs and yields the desired outputs, including the onset/offset and the mean of segments specified in the nodes of the learned constraint graph \cite{fotoohinasab2021graph}. Here, the model architecture is comprised of two stages: training and detection. The training step tries to heuristically find an optimum graph structure by which the label errors in the training set are minimized (the block named ``Graph Learning Algorithm'' in Figure \ref{fig:System_model}). The detection step then extracts the R-peaks in the raw ECG record constrained to the graph learned in the previous step (the block named ``Changepoint Detection Model'' in Figure \ref{fig:System_model}). 

\begin{figure}[t!]
      \centering
      \includegraphics[height=1.0\textheight,width=1.0\linewidth,keepaspectratio]{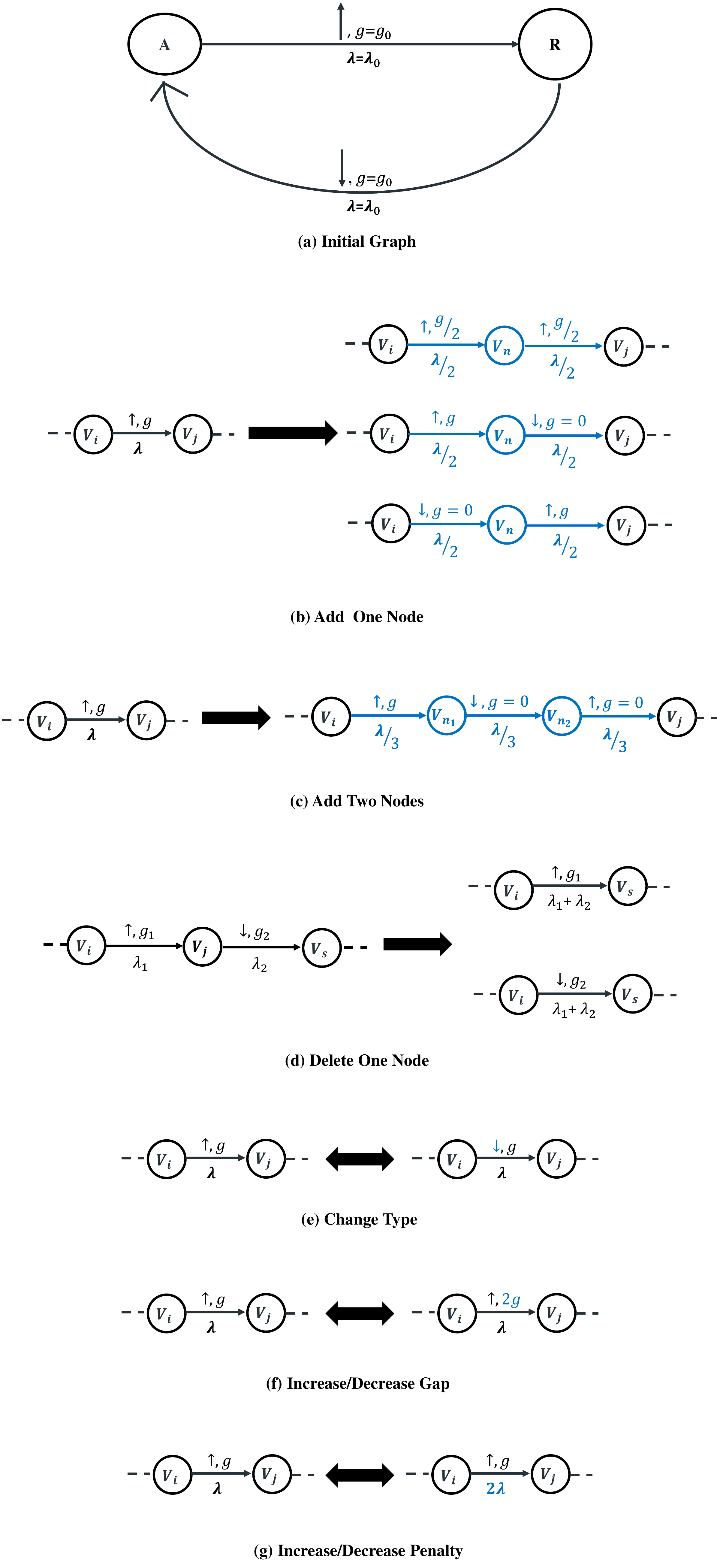}
      \caption{\textbf{(a:)} The initial constraint graph structure with two nodes labeled as $A$ and $R$, representing an alternative segment and the R-peak segment, respectively, in a cycle.
      \textbf{(b--g:)} Some of the applied graph editing candidates related to the edge $(V_i, V_j)$ with an up change.}
      \vspace{-15pt}
      \label{fig:Graph_Editing_Candidates}
\end{figure}

The novelty of this new structure lies in the training step, which is comparable to the previous model in Section \ref{sec: Manual graph definition}. The main idea of the training step is to automatically discover the desired topology of the constraint graph $G$ and the information about the edges from the data. As described in Section \ref{sec: Graph Constrained Changepoint Detection Model}, each edge contains the following information: (1) the expected up/down change in the segment means, (2) the least amplitude gap between the means of two states, and (3) a non-negative penalty imposed by the edge transition. Suppose that the initial graph for each record is denoted as $G_0 = (V_0, E_0)$, where $V_0$ and $E_0$ are the corresponding graph node and edge sets, respectively. Each node in the $V_0$ set represents initial hidden states in the model. Each edge in the $E_0$ set represents a transition between two consecutive hidden states (i.e., a changepoint $e$ from the source $\underline v_e$ to the target $\overline v_e$ in section \ref{sec: Graph Constrained Changepoint Detection Model}) and also contains initial values for parameters of $t_0$, $g_0$, and $\lambda_0$, which are the initial type, the initial gap between two states, and the initial penalty, respectively. Figure \ref{fig:Graph_Editing_Candidates}a shows the simple initial graph used for the optimization process. It should be noted that the initial edge information was chosen based on the overall results obtained from the manual definition of the constraint graph.   

A sketch of the proposed graph learning algorithm is summarised as Algorithm \ref{alg:training_alg}. The greedy graph search algorithm starts with the initial graph $G_0$ and iteratively optimizes the graph structure and edge parameters to find a graph that maximizes the accuracy regarding the provided labels. At the $t$-th iteration, the function $Find\char`_Graph\char`_Candidates()$ finds the graph candidate set $G^c_t$ using the editing candidates for each edge of the output graph from the previous iteration $G_{t-1}$. In this study, the algorithm considers 11 editing candidates per edge to optimize the graph topology and the three edge parameters. For example, in the iteration $t$, if the parent graph (i.e., $G_{t-1}$) has two edges, the graph candidate set $G^c_t$ will have no more than 22 members $|G^c_t|\leq 22$. 
These editing candidates include three types of adding a node, two types of deleting a node, one type of adding two nodes, changing the type of the abrupt change, and increasing or decreasing the penalty and gap corresponding to an edge.  We believe all morphological patterns of the ECG waves can be constructed using these editing candidates. Figure \ref{fig:Graph_Editing_Candidates} illustrates the graph editing candidates related to the edge $(V_i, V_j)$ with an up change.

\subsection{Computational Complexity}
\label{sec:Time Complexity}

As can be seen in Algorithm~\ref{alg:training_alg}, the time complexity of the GCCD algorithm is theoretically proportional to the number of graph candidates at each iteration (Line \ref{alg:for}) and the number of required iterations to achieve an optimum graph with minimum label errors (Line~\ref{alg:while}). 
There are three main aspects that characterize the time complexity of the algorithm:
\begin{itemize}
\item Given a record with $n$ data samples and a graph candidate $\hat{G}$ with $V$ vertices and $E$ edges, the time complexity to detect R-peaks (Lines~\ref{alg:10}--\ref{alg:if}) is $S = O(En^2)$ in the worst case (pathological simulated data) and $S = O(En\log n)$ in the average case (typical in real data). Also note that since we consider only graphs with a single circular path, $E=O(V)$, and the time complexity is further reduced to $O(Vn \log n)$ (for average case/non-pathological data).
\item Considering $C$ graph edit candidates in the iteration $t$, the time complexity to compute all the models $\hat{G}\in\{0,1,\dots,|G^c_t|=C\}$ is $O(SC)$ (where $S$ is the time complexity of solving for optimal model parameters given a single graph). It should be noted that the number of graph candidates in the iteration $t$ depends on $G_{t-1}$, which is the graph from the previous iteration (Line~\ref{alg:for}). The time complexity to compute the label error given $L$ labels is $O(CL)$, which can be effectively ignored from the overall time complexity as this task is fast.
\item Finally, iterating over $T$ iterations to obtain the graph with the minimum label error (Line~\ref{alg:while}) causes the overall time complexity of the algorithm to be $O(SCT)$, where $S$ is the time to solve for a single graph, and $C$ is the number of edit candidates considered in each iteration.
\end{itemize}

\begin{algorithm} \caption{Greedy Graph Learning}
\begin{algorithmic}[1]
\Input{data, labels, initial graph structure $G_0$}
\State $t \gets 0$
\State $Best\_Cost \gets \inf$
\State $\hat{E}_t \gets label\_error(G_t)$
\While{$\hat{E}_t < Best\_Cost$} 
\label{alg:while}
\State $Best\_Cost \gets \hat{E}_t$ 
\State $t \gets t+1$ 
\State \vspace{-.6cm} \begin{align*} 
 G^c_t \gets & Find\_Graph\_Candidates(G_{t-1}), 
 \\ & Based~on~ Figure~ \ref{fig:Graph_Editing_Candidates}
 \end{align*}
\vspace{-.6cm}
\State $\hat{E}_t \gets Best\_Cost$ 
\ForEach {$\hat{G}$ \textbf{in} $G^c_t$}
\label{alg:for}
\State $\hat{E} \gets label\_error(\hat{G})$
\label{alg:10}
\If {$\hat{E} < \hat{E}_t$}
    \State $\hat{G}_t \gets \hat{G}$
    \State $\hat{E}_t \gets \hat{E}$
\EndIf
\label{alg:if}
\EndFor
\EndWhile
\Output{constraint graph $G_t$}
\end{algorithmic}
 \label{alg:training_alg}
\end{algorithm}

\section{Experimental Studies}
\label{sec:Experimental Studies}
\subsection{Dataset}

We applied the well-known MIT-BIH Arrhythmia (MIT-BIH-AR) database to evaluate the GCCD model. This database contains 48 ECG recordings taken from 47 subjects. Each record's duration is 30 min, and each recording is sampled at 360 Hz with a resolution of 200 samples over a 10 mV range \cite{moody2001impact, goldberger2000physiobank}. Each recording consists of two ambulatory ECG channels from the modified lead II (MLII) and one of the leads V1, V2, V4, or V5. In this study, all 48 records with one MLII or V5 lead were used to evaluate the algorithm. The database has been annotated with both RR intervals and heartbeat class information by two or more expert cardiologists independently.

\subsection{Results and Discussion}
\label{sec:Results and Discussion}

This section presents a comprehensive discussion of the results obtained by the proposed model and a detailed comparison between the manually defined graphs and the learned graphs. We also provide some suggestions for the future development of the GCCD model. 

\begin{figure}[t]
      \centering
      \includegraphics[height=1.0\textheight,width=1.0\linewidth,keepaspectratio]{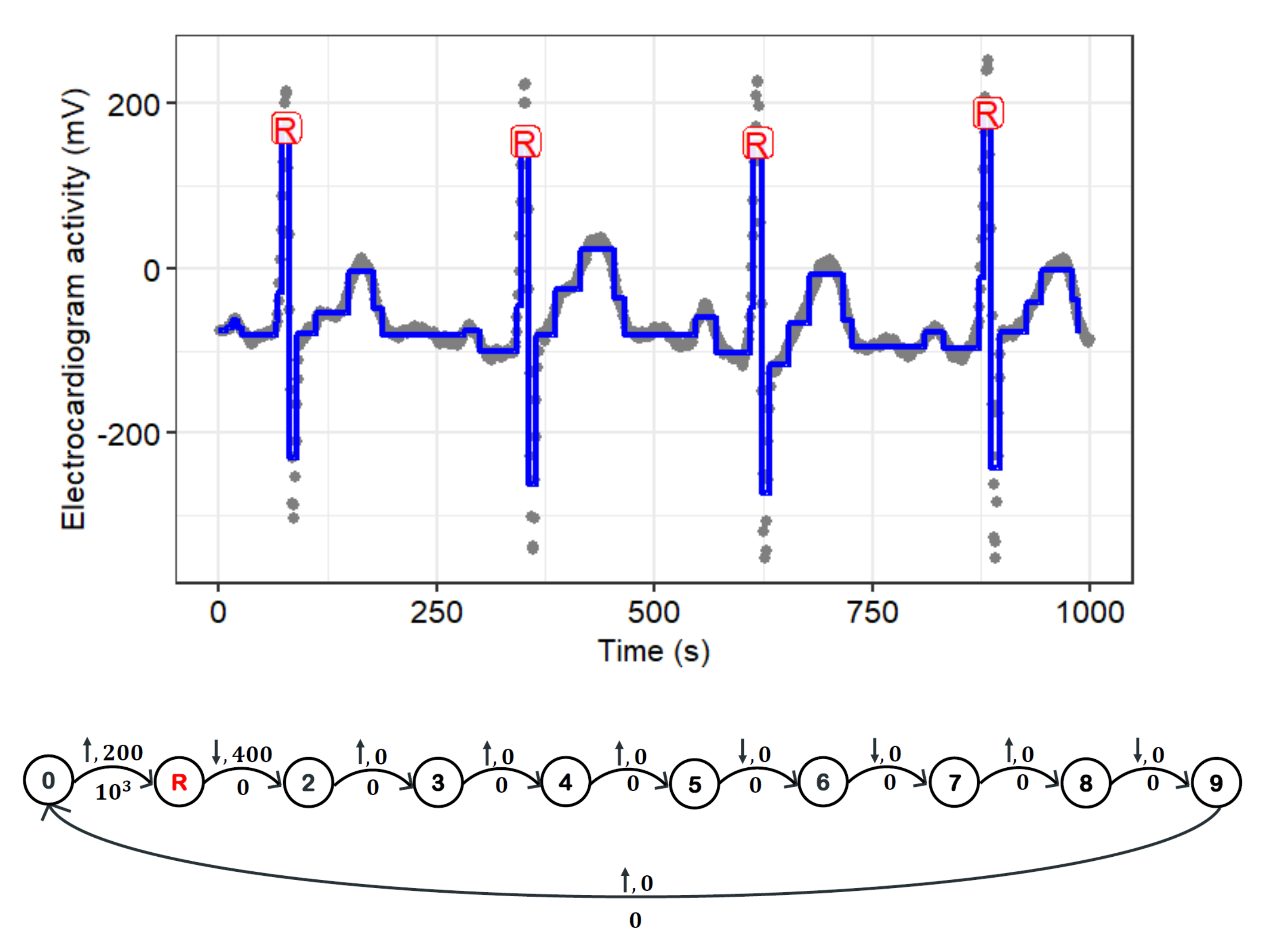}
      \caption{Demonstration of R-peak detection using the proposed model on Record 230 of the MIT-BIH-AR dataset.
    \textbf{(Top:)} The proposed model represents Record 230 as piece-wise locally stationary segments (blue lines). Extracted R-peak positions are marked with a red ``R."
    \textbf{(Bottom:)} The graph structure for the proposed model. The constraint graph has a vertex for each state including state ``R" for the R-wave. Below each edge $e$ we show the penalty $\lambda_e$, which is either a constant $\lambda>0$ or $0$; above we show the constants $\delta,\gamma$ in the constraint function $g_e(m_i,m_{i+1})= \delta(m_i - m_{i+1})+\gamma\leq 0$ , where $\delta=1$ for a non-decreasing change (shown with $\uparrow$), $\delta=-1$ for a non-increasing change (shown with $\downarrow$), and $\gamma \geq 0$ is the minimum magnitude of change.}
      \vspace{-15pt}
      \label{fig:Manual_230}
   \end{figure}

Figure \ref{fig:Manual_230} illustrates an example of the model's performance with a manually defined constraint graph in the R-peak detection task for a window of Record 230 of the MIT-BIH-AR dataset. However, as mentioned in Section \ref{sec: Manual graph definition}, the performance of the model using manually defined graphs depends on an expert with prior knowledge. Furthermore, manual annotation by an expert is time consuming and expensive. To address this issue, we proposed a new graph learning algorithm that searches for a locally optimal constraint graph using a greedy scheme on the labeled ECG data. Regarding the various morphological patterns for the ECG signal, the proposed graph learning algorithm can relax the model from the manual definition of the constraint graph for each record.

   \begin{figure*}[htb]
      \centering
      \includegraphics[width=\linewidth,height=0.85\textheight,keepaspectratio]{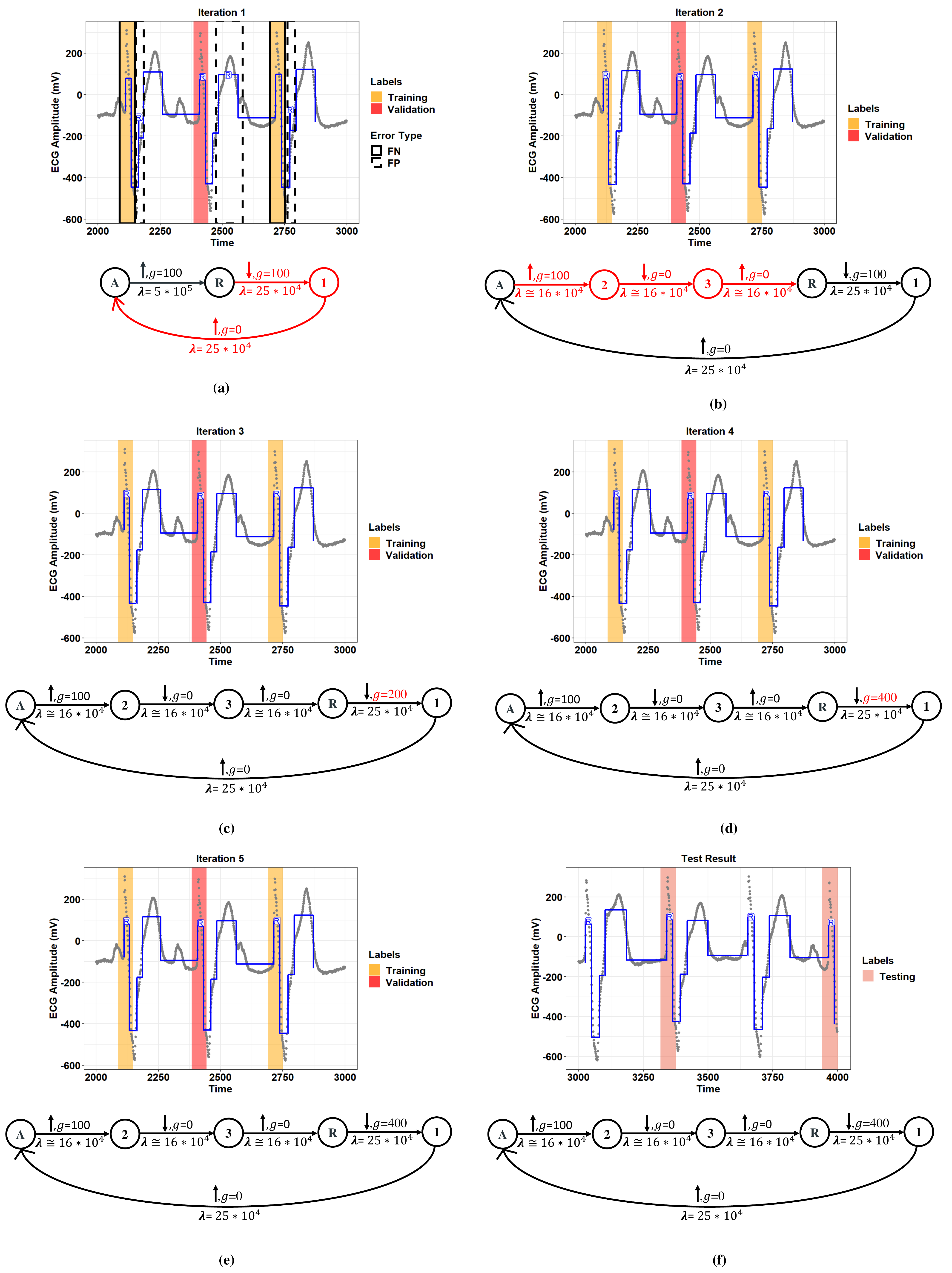}
      \caption{Demonstration of constraint graph optimization using the proposed graph learning algorithm for Record 107 of the MIT-BIH-AR dataset.
     \textbf{(a--e, top:)} Extracted R-peak positions given the learned constraint graph in each learning iteration. The red and orange coverage bands show the used labels in the training procedure, including training and validation sets. The blue lines represent separate states at the model output over the raw ECG signal (gray points). 
     \textbf{(a--e, bottom:)} The learned constraint graph in each learning iteration. For each edge, $\lambda$ is the penalty, $g$ is the gap (the minimum magnitude of change), and the up/down arrow shows the type of change. The part of the graph that is modified from the previous iteration is shown in red. 
      \textbf{(f:)} Testing results for the final learned constraint graph using a new window of data. The pink coverage bands show the labels in the testing set.}
      \vspace{-15pt}
      \label{fig:Learning_107}
   \end{figure*}

\begin{figure*}[htb]
      \centering
      \includegraphics[width=\linewidth,height=0.91\textheight,keepaspectratio]{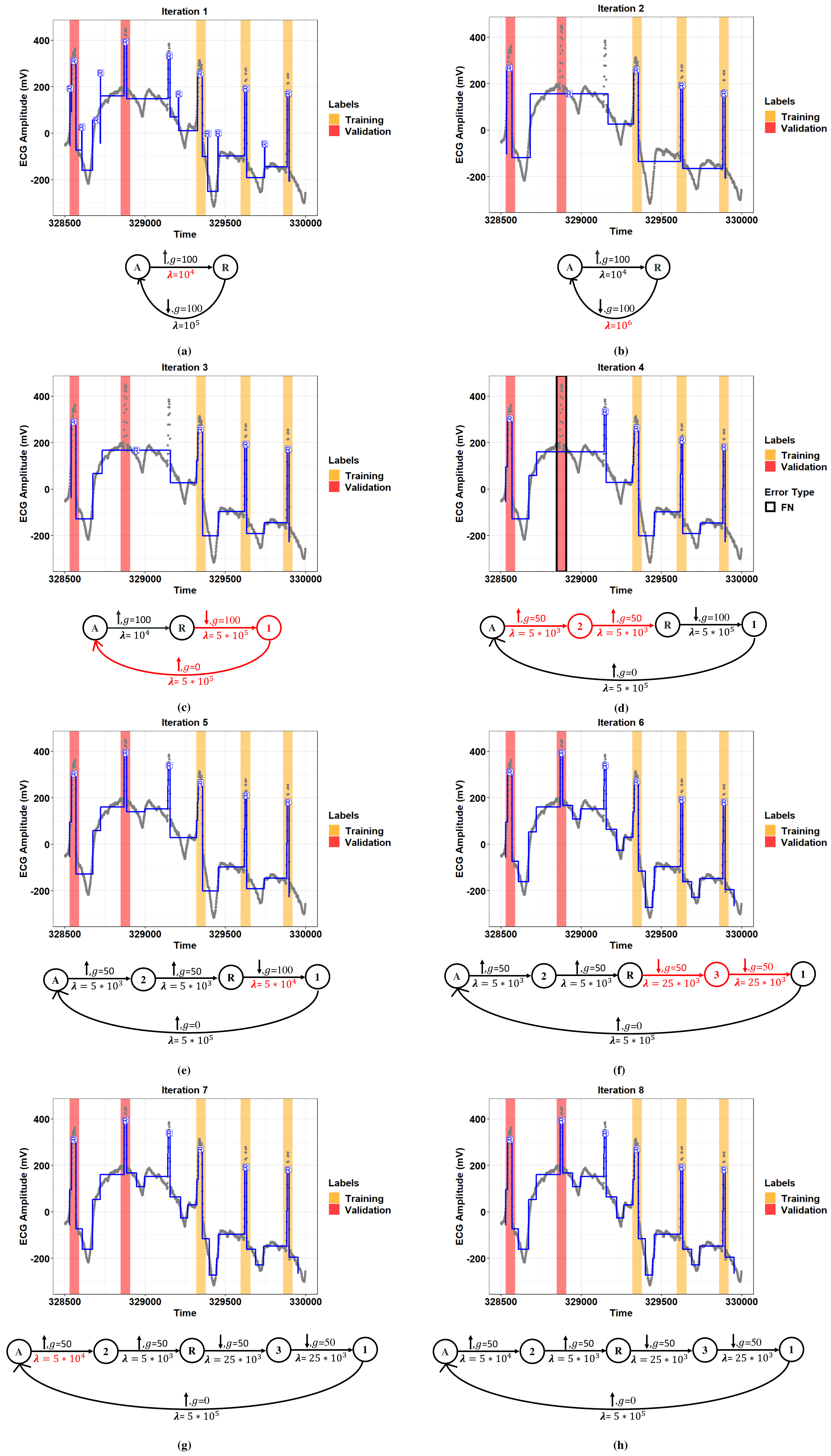}
      \caption{Detection of R-peak positions for Record 219 of the MIT-BIH-AR dataset using the proposed graph learning algorithm.
     \textbf{(a--h, top:)} Extracted R-peaks given the constraint graph structure learned in each iteration. The orange and red coverage bands present training and validation labels, respectively. The blue line demonstrates locally stationary segments at the model output, and the gray points also show the raw ECG signal at the model input. \textbf{(a--h, bottom:)} The constraint graph learned in each iteration. The red part of the constraint graph represents the selected graph editing candidate in each iteration.}
      \vspace{-15pt}
      \label{fig:Learning_219}
   \end{figure*}

We adopted the intra-patient paradigm to train the constraint graph to address the intra-patient variation in ECG morphologies. Thus, the training and testing sets were generated by randomly splitting the intra-samples for each record with an approximate ratio of $3:1$. We used a $k$-fold cross-validation approach to evaluate the model performance with a $k$ size of $5$. More specifically, we divided the intra-sample data into $k=5$ folds so that each trial used four folds to train the model and one fold for validation. 

\begin{figure*}[htb]
      \centering
      \includegraphics[width=\linewidth,height=\textheight,keepaspectratio]{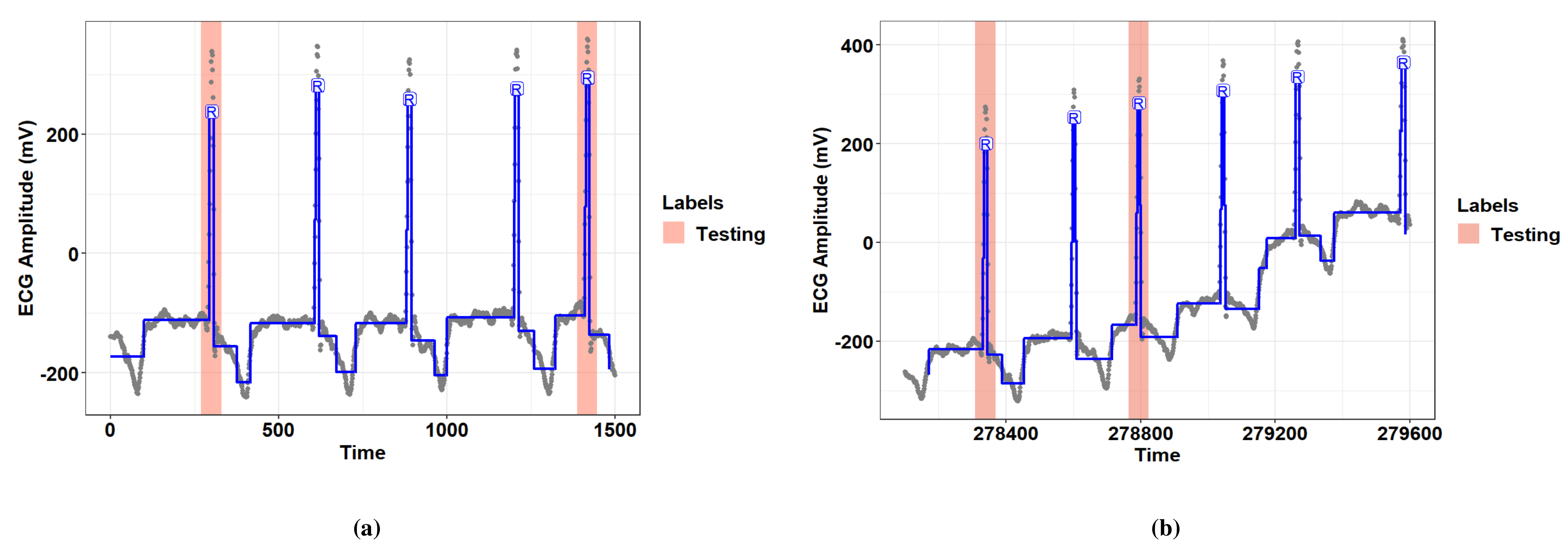}
      \caption{Test result for Record 219 for two different windows of time. The pink coverage bands represent the labels in the testing set, and the blue lines demonstrate model output.}
      \vspace{-15pt}
      \label{fig:Test_219}
   \end{figure*}

\begin{figure}[t!]
     \begin{subfigure}[t]{0.5\textwidth}
       \centering\includegraphics[width=\textwidth]{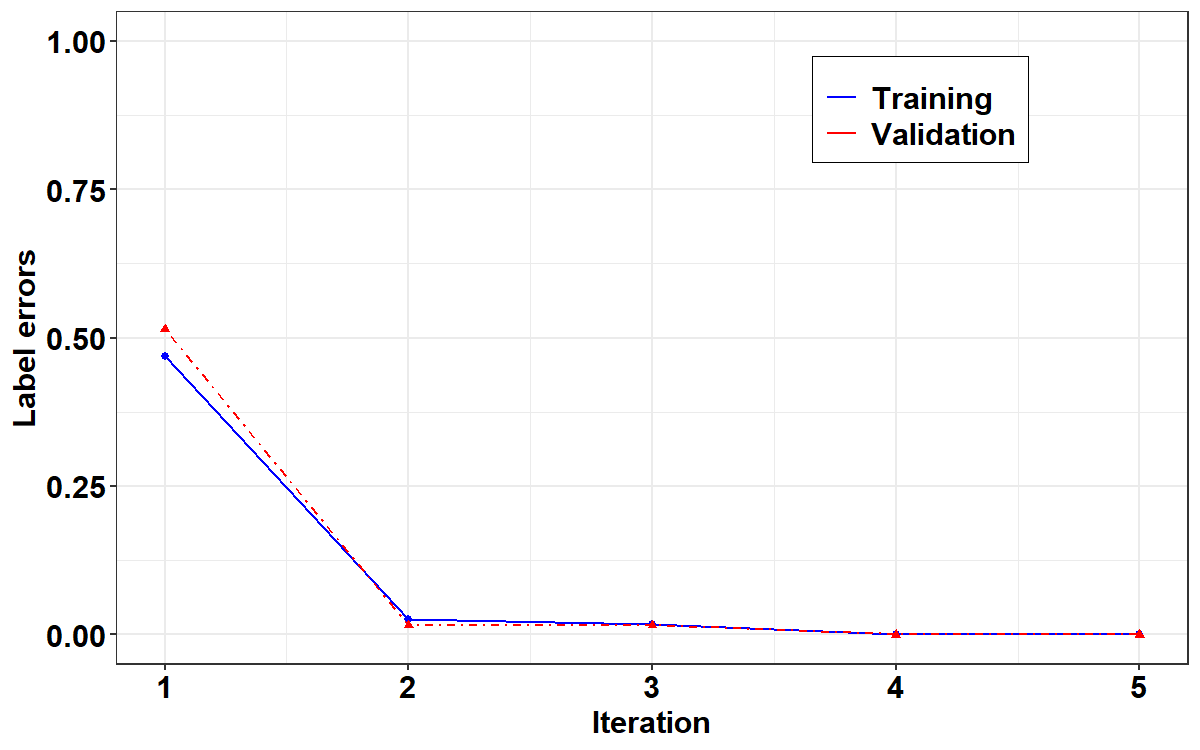}
       \caption{}
     \end{subfigure}
     \begin{subfigure}[t]{0.5\textwidth}
       \centering\includegraphics[width=\textwidth]{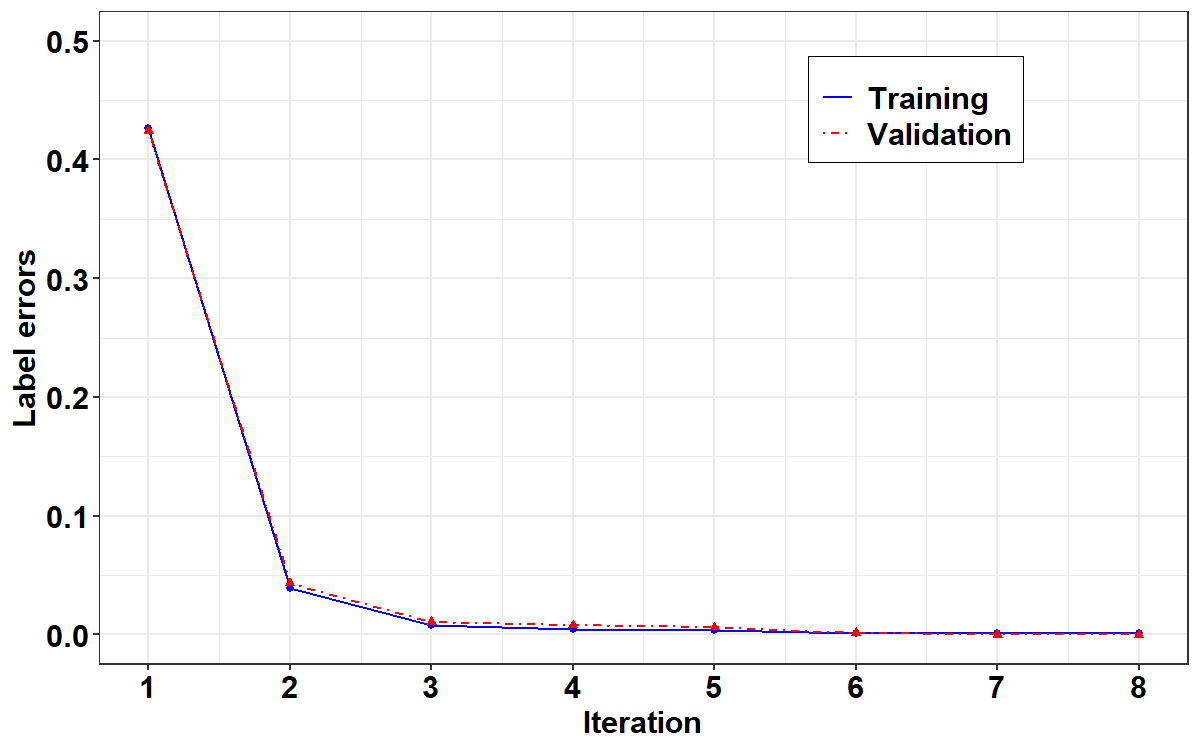}
       \caption{}
     \end{subfigure}
     \caption{Training progress for \textbf{(a)} Record 107 and \textbf{(b)} Record 219 of the MIT-BIH-AR dataset.}
      \label{fig:TrainingProgress_107_219}
   \end{figure}

Figures \ref{fig:Learning_107} and \ref{fig:Learning_219} show representative examples of the R-peak detection task performed by the model integrated with the graph learning algorithm for two records from the MIT-BIH-AR database. These figures illustrate how the proposed graph learning algorithm iteratively edits the graph structure to yield a model with maximum accuracy in detecting R-peaks. We initialized the constraint graphs using the graph structure in Figure \ref{fig:Graph_Editing_Candidates}a with the initial values of $g_0=100$ and $\lambda_0=5\times10^5$ for Record 107 and $g_0=100$ and $\lambda_0=10^5$ for Record 219. It should be noted that the initial edge information was assigned based on the overall results derived from the manually defined graphs in all experiments. However, graph candidates \ref{fig:Graph_Editing_Candidates}f and \ref{fig:Graph_Editing_Candidates}g can adjust the parameters $g$ and $\lambda$ for the optimum values. For these two examples, we chose the initial edge information so that all the training steps could be completely displayed. Label errors are omitted from Figures \ref{fig:Learning_219}a--\ref{fig:Learning_219}c to reduce clutter in the figures. The red part of the graph structure in each iteration presents the chosen editing candidate in the current iteration over the graph in the previous iteration. More interestingly, Figure \ref{fig:Learning_219} demonstrates the model's capability to detect R-peaks in the presence of a baseline wandering artifact, which is a typical artifact in the ECG signal. Baseline wandering can change the shape of the QRS complex and thereby causes incorrect detection of the R-peak. The performance of the Pan and Tompkins \cite{PanTompkins1985} algorithm, algorithms derived from the ECG signal slope \cite{sabzevari2018ultra}, and methods based on wavelet transform are highly dependent on the removal of this artifact. Figure \ref{fig:Test_219} shows the test result for this record over two different time windows of data. Figure \ref{fig:TrainingProgress_107_219} illustrates the training progress for these two records, where the Y-axis shows the sum of false negative and false positive error rates. Indeed, the training progress curve reflects the number of label errors produced by the model in each iteration given the provided labels for the training and validation sets. It is worth mentioning that the proposed graph learning algorithm avoids possible overfitting issues as it tries to extract the morphology of the ECG signal that contains multiple various morphological patterns.

 \begin{figure}[t!]
      \centering
      \includegraphics[height=1.0\textheight,width=1.0\linewidth,keepaspectratio]{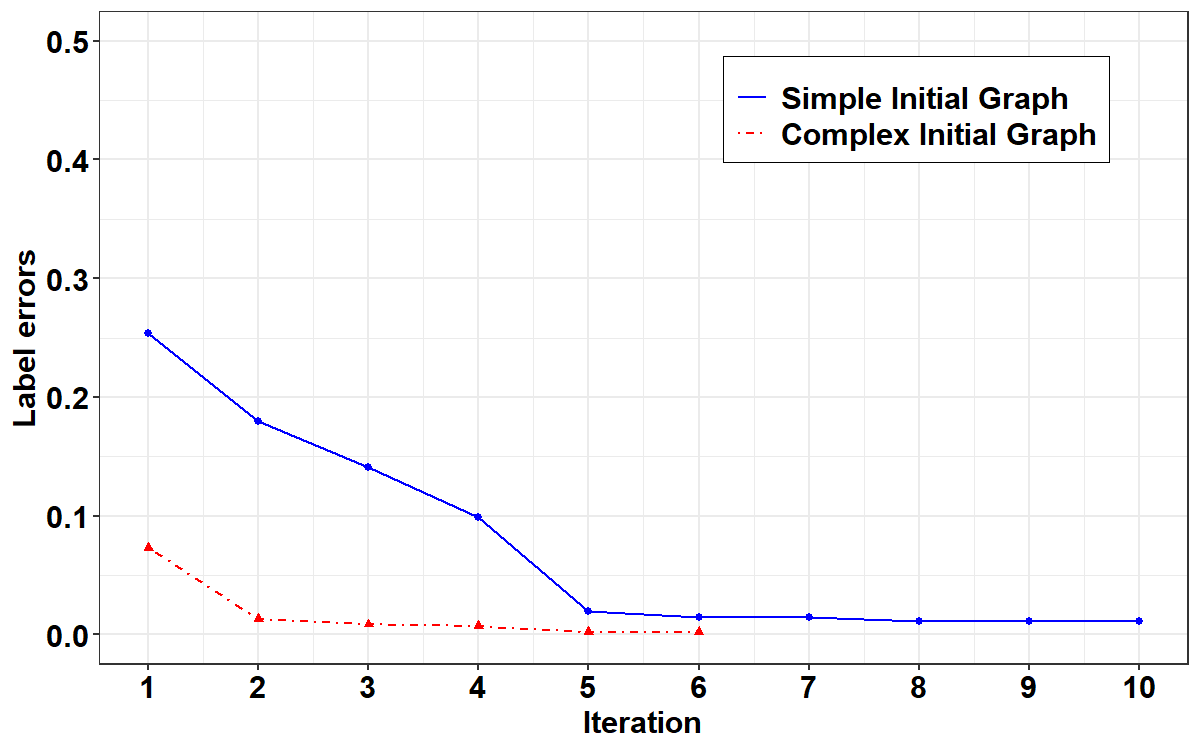}
      \caption{Comparison of the training progress initialized with two simple and complex initial graph structures for the record 106 in the MIT-BIH-AR dataset.}
      \vspace{-15pt}
      \label{fig:TrainingProgress_106}
   \end{figure}
   
  \begin{figure}[htb]
      \centering
      \includegraphics[height=1.0\textheight,width=1.0\linewidth,keepaspectratio]{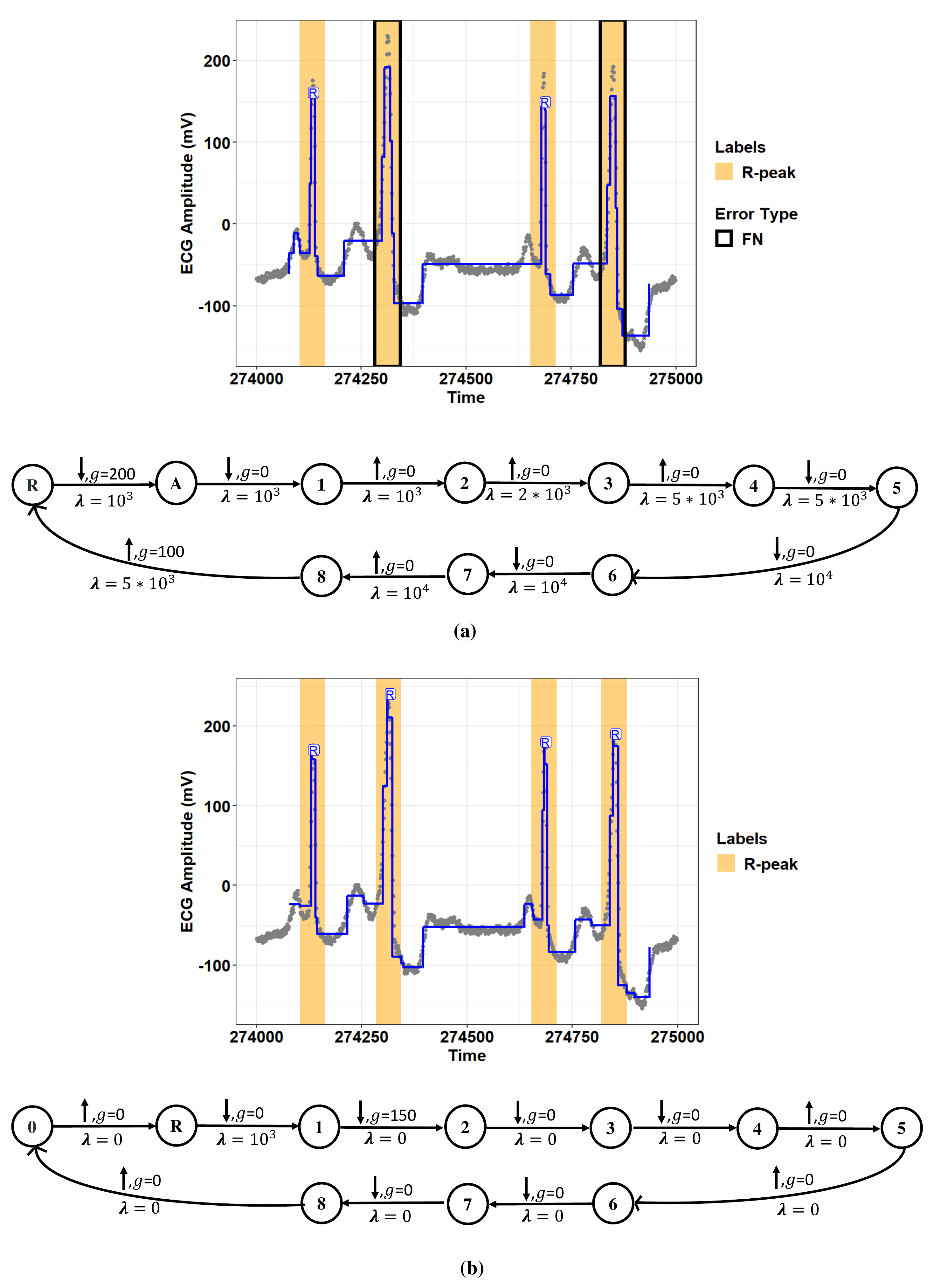}
      \caption{Comparison of the trained models for Record 106 of the MIT-BIH-AR dataset inititalized with \textbf{(a)} a simple graph structure and \textbf{(b)} a complex graph structure.}
      \vspace{-5pt}
      \label{fig:Comparison_106}
   \end{figure}   
   
The proposed graph learning algorithm employs a greedy search scheme to select the best performing graph in terms of detection accuracy (see  Section \ref{sec:Graph Learning}). Therefore, the performance of the algorithm depends heavily on the initial graph structure and will likely lead to local optima. Figure \ref{fig:TrainingProgress_106} compares the training progress for Record 106 of the MIT-BIH-AR database initialized with the two simple (see Fig. \ref{fig:Graph_Editing_Candidates}a) and complex graph structures (i.e., a graph with eight nodes representing the morphology of a normal ECG signal). Figure \ref{fig:Comparison_106} also presents a comparison of the final selected graphs and their performances for a window of this record. As these figures show, the model initialized with the complex graph structure can achieve higher accuracy (i.e., a lower number of label errors) in a lower number of iterations than the model initialized with the simple graph structure. 

\begin{figure}[t!]
      \centering
\includegraphics[height=1.0\textheight,width=1.0\linewidth,keepaspectratio]{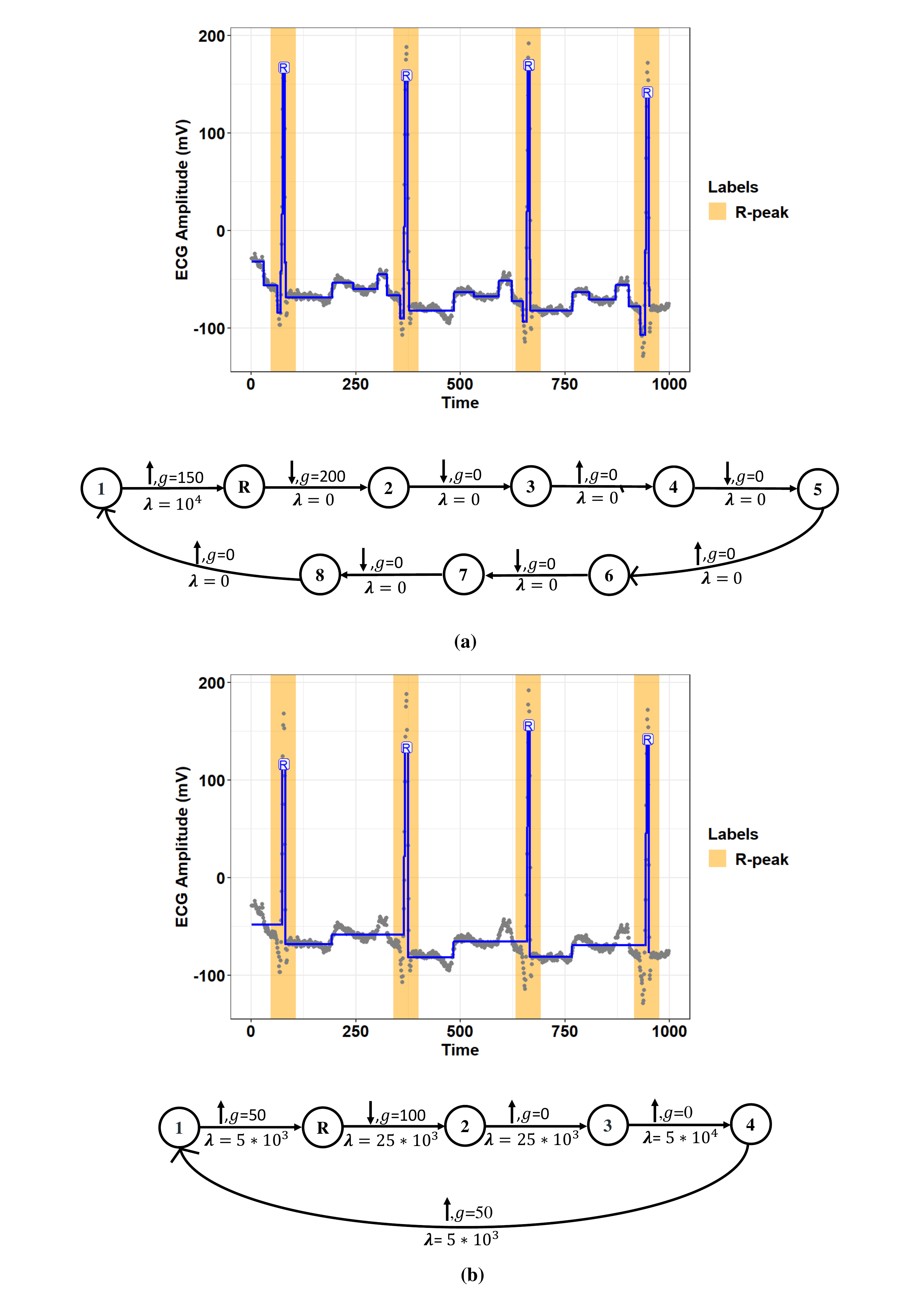}
      \caption{Comparison of the constraint graph structures \textbf{(a)} defined manually and \textbf{(b)} learned using the proposed graph learning algorithm for Record 100 of the MIT-BIH-AR dataset.}
      \vspace{-15pt}
      \label{fig:Comparison_patient_100}
   \end{figure}

\begin{figure}[htb]
      \centering
      \includegraphics[height=1.0\textheight,width=1.0\linewidth,keepaspectratio]{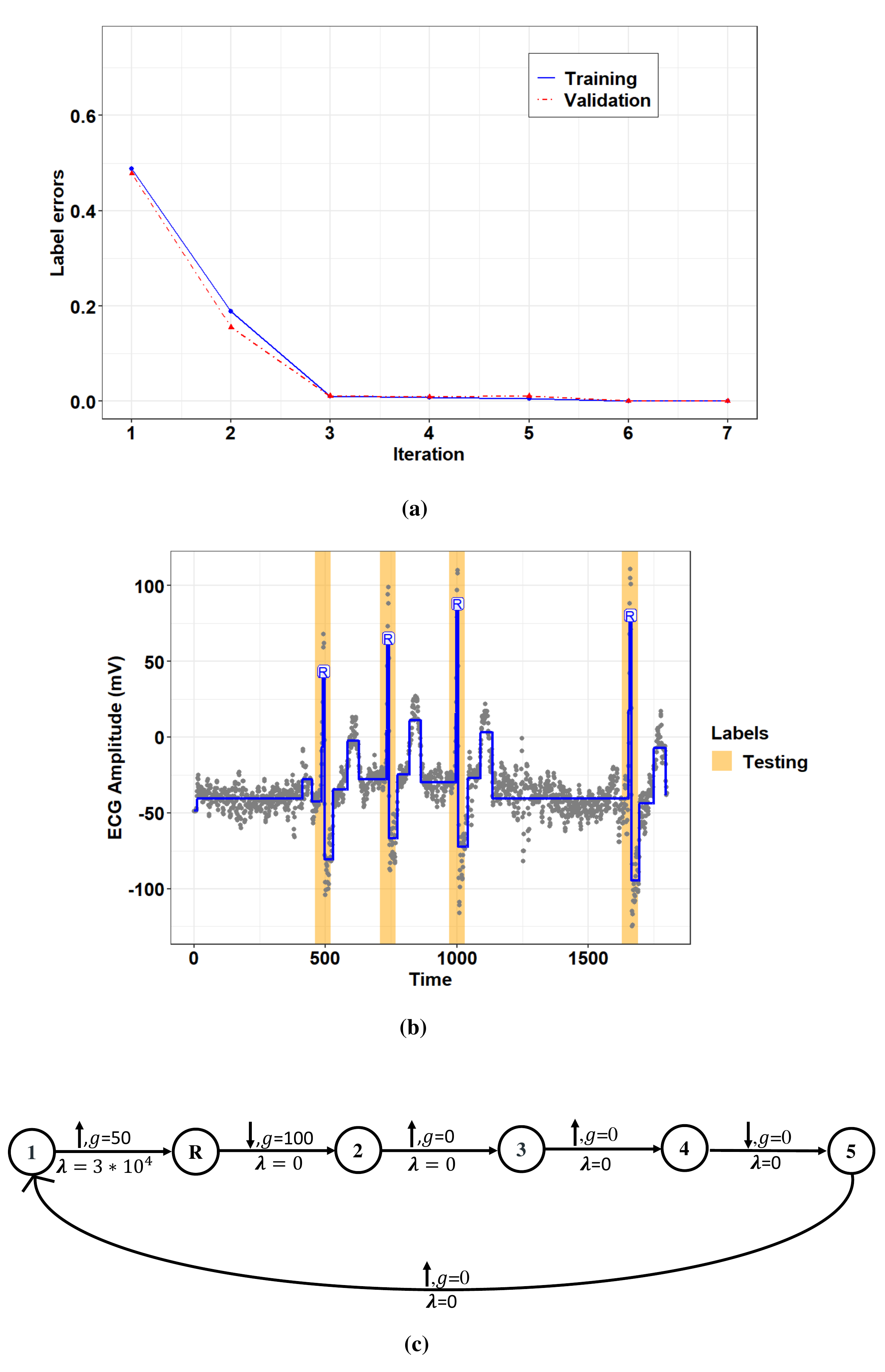}
      \caption{The model's performance for Record 230 of the MIT-BIH-AR dataset. \textbf{(a)} The training progress, \textbf{(b)} extracted R-peaks for a window of this record, and \textbf{(c)} the constraint graph structure.}
      \vspace{-15pt}
      \label{fig:patient_232}
   \end{figure}

The investigation of the experimental results shows that the greedy graph search algorithm can achieve optimal performance for the model trained with the manually defined graphs, although its performance is affected by the initial graph. We noticed that for most of the records from the MIT-BIH-AR database, the learned graphs could reach the performance of the manually defined graphs but with different graph structures. This means that the GCCD model can obtain global optima using various initialization structures, which will likely lead to different final graph structures. Figure \ref{fig:Comparison_patient_100} compares the constraint graph structures defined manually vs. those learned automatically using the initial graph structure in Figure \ref{fig:Graph_Editing_Candidates}a for Record 100 of the MIT-BIH-AR dataset. As shown in this figure, the manually defined graph and the learned graph both achieved the optimal performance but with different graph structures. We also noticed that for some records from the MIT-BIH-AR database, the graph learning algorithm chose the same structure as the manually defined graph structure. Figure \ref{fig:patient_232} shows the model performance using the graph learning algorithm for Record 232 from the MIT-BIH-AR dataset, for which the manually defined constraint graph and the learned graph had the same structures.

Different metrics were adopted to evaluate the performance of the proposed model with both the manual and learning-based graph designs. These metrics included sensitivity ($Sen$), positive predictivity rate ($PPR$), and detection error rate ($DER$), which are calculated by:

\begin{align}
\label{eq:sen}
\textit{Sen} (\%) =  \frac{TP}{TP+FN}\times{100}\\[10pt]
\label{eq:PPR}
\textit{PPR}  (\%) =  \frac{TP}{TP+FP}\times{100}\\[10pt]
\label{eq:DER}
\textit{DER} (\%) =  \frac{FN+FP}{TP+FN}\times{100}
\end{align}
where $TP$ is true positives, $FP$ is false positives, $FN$ is false negatives, and $TN$ is true negatives. Table \ref{tab:Comparison_of_methods_MIT} presents the performance of the proposed model regarding both the manually defined and learning graphs against other state-of-the-art methods for R-peak detection (QRS complex). As shown in the table, the proposed algorithm achieved \textit{Sen} = \%99.76, \textit{PPR}  = \%99.68, and \textit{DER} = 0.55 for the manual definition of the constraint graph and \textit{Sen} = \%99.64, \textit{PPR}  = \%99.71, and \textit{DER} = 0.19 for the learning constraint graph using the MIT-BIH-AR database. Note that the model constrained to the manually defined graphs outperformed the model combined with the graph learning algorithm because in the latter, the model's performance was largely dependent on the initial graph structure. 
\begin{table}[ht]
\caption{Comparison of the performance of several R-peak detection methods using the MIT-BIH-AR database}
 \centering{
\label{tab:Comparison_of_methods_MIT}
\resizebox{1.\linewidth}{!}{ 
\begin{tabular}{cccc}
\toprule
\textbf{Method} & \textbf{\textit{Sen} (\%)} & \textbf{\textit{PPR}  (\%)} & \textbf{\textit{DER} (\%)}\\ 
\midrule
\texttt Park et al. \cite{park2017r} & 99.93 & 99.91 & 0.163\\
\texttt Farashi \cite{farashi2016multiresolution} & 99.75 & 99.85 & 0.40 \\
\texttt Sharma and Sunkaria \cite{sharma2017qrs} & 99.50 & 99.56 & 0.93\\
\texttt Castells-Rufas and Carrabina \cite{castells2015simple} & 99.43 & 99.67 & 0.88\\
\textbf {GCCD Model with Manual Definition of the Constraint Graph} & \textbf {99.76} &  \textbf {99.68} & \textbf {0.55}\\
\textbf {GCCD Model with Learning of the Constraint Graph} & \textbf {99.64} &  \textbf {99.71} & \textbf {0.19}\\
\hline
\end{tabular}}}
\end{table}

ECG recordings in the MIT-BIH-AR database were chosen to challenge the R-peak detection task because they represent a wide variety of QRS morphologies with real-world variability. Our proposed model yielded outstanding results when detecting R-peaks in these tricky records. Records 103, 104, 105, 108, 111, 112, 116, 200, 201, 203, 205, 208, 210, 217, 219, 222, and 228 are comprised of abrupt changes in ECG morphology, and they are severely affected by noise and artifacts. Figure \ref{fig:Learning_219} shows the capability of the model to detect R-peaks in the presence of baseline wandering noise. We re-emphasize that these comparable results were obtained without applying any preprocessing operations, as opposed to other methods in the literature. Records 108, 113, 117, 201, 202, 203, 213, 219, 222, 223, 231, and 232 contain many peaks with unusual amplitudes. Small-amplitude R-peaks or high-amplitude P- and T-peaks embedded in high-amplitude QRS complexes can lead to high FN and FP errors in the R-peak detection task. As a representative example, Figure \ref{fig:Learning_107} illustrates the efficiency of the GCCD model in R-peak detection for Record 117, which contains many beats with high-amplitude T-peaks. 

The experimental results obtained using the proposed model justify changepoint detection models as a potential approach to extract ECG fiducial points. In this study, we demonstrated the capability of the GCCD model in locating R-peaks within various morphological patterns of ECG. The proposed greedy graph search algorithm can potentially detect ECG waves other than the R wave (i.e., P, Q, S, and T waves) by considering corresponding prior knowledge of the graph editing candidates. We noticed that in Records 114, 200, 203, 207, and 210, the \textit{Sen} and \textit{PPR} values were less than 99\%. These records contain multiple different morphological patterns, including negative QRS complexes, and Records 200 and 203 have several QRS complexes with ventricular arrhythmias. The constraint graph for these records involves learning a graph with more than one optimum graph path. Learning a multi-path constraint graph is also required to detect all ECG waves due to the various morphological patterns of each wave incorporated into the graph. The other point that should be considered here is that the GCCD model estimates the ECG signal using a Gaussian function. A modified model with a multi-Gaussian fitting method can drastically improve the ECG-related changepoint detection task.

Future work should focus on developing the proposed model with a multi-Gaussian fitting and a multi-path graph learning algorithm. Incorporating these modifications into the proposed model could provide a promising platform for evolving new graph-based tools to detect and classify heart arrhythmias. A multi-path graph learning algorithm could reveal the morphology of the ECG signal (time duration, amplitude, and direction of each wave) in each cardiac cycle. Subsequently, new graph-based features could be extracted from the constraint graph path for an ECG cycle to classify heartbeats. 

\section{Conclusion}
\label{sec:CONCLUSIONS and DISCUSSION}

The accurate delineation of R-peaks in the ECG signal plays a crucial role in most automated ECG analysis tools. This paper proposed a novel graphical model based on changepoint detection techniques for detecting R-peaks within a non-stationary ECG signal. The proposed model was highly successful at detecting R-peaks in noisy ECG data without applying any preprocessing steps. To our knowledge, this is the first time that a changepoint detection model has been applied for ECG fiducial points detection. In this new framework, prior biological knowledge about the expected sequences of changes was incorporated into the model using a graph. We defined the constraint graph manually and automatically using a proposed greedy graph search algorithm. Using the proposed graph learning algorithm, the initial graph structure can develop into a structure containing edge parameters with maximum detection accuracy for a record. The experimental results provided in this paper demonstrate that the GCCD model can be a promising approach for detecting ECG waves and developing new graph-based tools for further ECG analysis. The proposed graphical model approach can be advanced by learning a multi-path constraint graph and fitting a multi-Gaussian curve model to the ECG signal, which should be considered in future studies.

\section*{ACKNOWLEDGMENT}
This material is based on work supported by the National Science Foundation under Grant Number 1657260. Research reported in this publication was supported by the National Institute on Minority Health and Health Disparities of the National Institutes of Health under Award Number U54MD012388.
\printbibliography

\end{document}